\documentclass[aps,prl,onecolumn,groupedaddress,14pt]{revtex4}
\usepackage[english]{babel}
\usepackage{mathtext,amssymb,amsmath,float}
\usepackage{graphics}
\usepackage{epsfig}

\begin{document}
\title{Transient sequences in a hypernetwork\\ generated by an adaptive network of spiking neurons}
\author{Oleg~V.~Maslennikov}
\email{olmaov@ipfran.ru}
\author{Dmitry S. Shchapin}
\author{Vladimir~I.~Nekorkin}
\affiliation{Institute of Applied Physics of RAS, Nizhny Novgorod, Russia}

\date{\today}
\begin{abstract}
We propose a model of an adaptive network of spiking neurons that gives rise to a hypernetwork of its dynamic states at the upper level of description. Left to itself, the network exhibits a sequence of transient clustering which relates to a traffic in the hypernetwork in the form of a random walk. Receiving inputs the system is able to generate reproducible sequences corresponding to stimulus-specific paths in the hypernetwork. We illustrate these basic notions by a simple network of discrete-time spiking neurons together with its FPGA realization and analyze their properties.
\end{abstract}

\maketitle

\section{Introduction}
The interest to multiagent systems with entangled structures and nontrivial dynamics known as complex networks is stimulated by their prevalence in nature, technical applications and society, and by the common mathematical language mainly based  on graph theory, statistical physics and nonlinear dynamics~\cite{newman2003structure,boccaletti2006complex,arenas2008synchronization}. The most promising application field of the complex network view is neuroscience~\cite{bullmore2009complex,rubinov2010complex,hagmann2008mapping,sporns2004organization,jirsa2004connectivity}. Revealing the architecture of structural links in neuronal structures  as well as functional relations between anatomically remote brain regions allows formulating basic principles of the organization of the central nervous system in terms of complex networks and nonlinear dynamics~\cite{sporns2010networks,rabinovich2006dynamical}. New findings in neuroscience require developing novel concepts in mathematical descriptions; for example, the hierarchical structure and activity in brain stimulates generalizations in  network science, e.g., temporal, multiplex, multilayer, multilevel networks~\cite{holme2012temporal,de2013mathematical,boccaletti2014structure,kivela2014multilayer}, and hypernetworks~\cite{johnson2013hypernetworks,johnson2016hypernetworks}.

The ever-changing synaptic couplings lead to studying adaptive or evolving dynamical networks~\cite{gorochowski2012evolving,belykh2014evolving,maslennikov2015evolving,maslennikovUFNadaptive}, impermanent nature of most perception and cognitive processes require shifting focus in nonlinear dynamics from attractors to transients~\cite{rabinovich2008transient,rabinovich2012principles}. The transient dynamics of spatiotemporal patterns and evolving architectures of synaptic links seem to be basic properties underlying most complex brain functions ~\cite{hutchison2013dynamic,petersen2015brain,mivsic2016regions}.
Many  studies of anatomical structures and physiological processes support the idea that cognitive processes relate to interactions among distributed neuronal populations and brain areas~\cite{park2013structural,sporns2014contributions}. It was hypothesized and experimentally shown that during each perception or cognitive act there emerges an assembly of spatially distributed neurons---the dynamic core---that operates in a coordinated manner~\cite{tononi1998consciousness,thompson2001radical,varela2001brainweb,le2003disentangling,edelman2003naturalizing,tononi2004information}. The neurons within the assembly preferentially communicate with each other giving rise to some integration activity. This process is highly transient which enables the neuronal network to rapidly change the activation pattern reflecting the multitasking property of brain and functional segregation~\cite{buonomano2009state,buzsaki2010neural,rigotti2013importance,stokes2013dynamic}. Uncovering biophysical and dynamical mechanisms underlying a rich functional repertoire of the brain circuit at the micro-, meso-, and macrolevels remains a nontrivial challenge for both experimentalists and theoreticians. Understanding  the principles of relating cognitive functions to coordinated activities of neuronal substrates is still in its infancy and needs to be described by an appropriate mathematical language. In this paper we propose a descriptive model for emerging complex behaviours in a hypernetwork---at the upper level with respect to a basic network  with co-evolving coupling structure and nodal dynamics. This approach is used to provide an  example of how a simple structured network of coupled oscillators gives rise to the traffic in the hypernetwork.

\section{Results}\label{subsec_hypernetwork}
\subsection{A hypernetwork generated by an adaptive network}
We propose a paradigmatic model based on the concept of hypernetworks~\cite{johnson2013hypernetworks,johnson2016hypernetworks}; this model captures how network dynamics and structure evolution can lead to the emergence of a new level of description which may be considered as an emergence of functions from evolving dynamical structures. The model gives a sketch, a mathematical image of how a small oscillatory network with co-evolving structure and dynamics generates a new -- functional -- structure at an upper level of description. Simple transient dynamics of activity clusters result in a special form of traffic in this new structure that we refer to as a hypernetwork~\cite{anokhin2016}.

\begin{figure}[H]
\centering\includegraphics[width=4.5in]{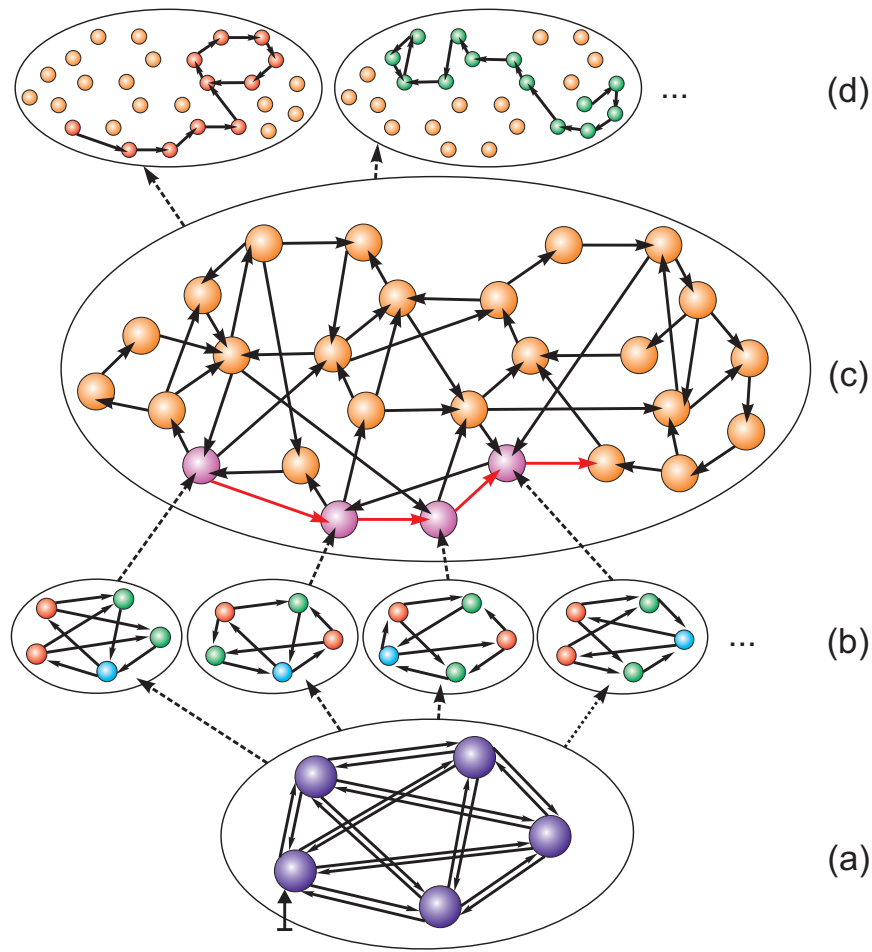}
\caption{The network at the bottom level (a) generates a family of relational simplices (b) that appear sequentially due to the co-evolution of structure and dynamics. The set of hypersimplices, or simplicial family, are connected to each other by directed links thus constituting a hypernetwork (c). Different paths in the hypernetwork (d) occur according to different stimuli (the vertical arrow) coming to the bottom level (a).   }
\label{fig_hypernetwork}
\end{figure}

The outline of the model is shown in Fig.~\ref{fig_hypernetwork}.  First we consider a neuronal population as a network of interacting nodes-oscillators coupled by directed links. This is the bottom level presented schematically in Fig.~\ref{fig_hypernetwork}(a). The nodal dynamics as well as the activity of links are governed by some deterministic evolutionary operators (for details, see \emph{Materials and Methods}). Depending on nodal dynamics the links become  active or inactive thus forming different structural patterns in the network. Physiologically this can mean that some synaptic links become strong while others tend to be weak.  A structural pattern results in a specific spatiotemporal activity in the form of synchronous or asynchronous clusters. Depending on the internal state as well as external conditions the network can adapt to different stimuli and reconfigure the structure of strong and weak couplings thus changing the spatiotemporal pattern. Different perception stimuli or cognitive tasks give rise to the activation of different distributed neuronal groups or to a different order of sequential activation of various brain areas. We relate this distributed neuronal group specified by the common task to a corresponding hypersimplex.  \emph{Hypersimplices} are ordered, or structured, sets of nodes with an explicit relation; in other words, they exist at a higher level of representation than network's nodes (see Fig.~\ref{fig_hypernetwork}(b)). In our model hypersimplices are specified by the coupling topology that leads to a certain cluster activity. By a \emph{cluster activity} we mean the form of collective behaviour exhibited by oscillators. For example, in Fig.~\ref{fig_hypernetwork}(b) nodes with a same colour denote one cluster, i.e., they tend to fire synchronously while other nodes are silent. Different colours relate to different clusters, i.e., these groups of oscillators do not fire simultaneously. The cluster activity which is a functional attribute of the network is directly specified by the current hypersimplex which is a structural attribute of the network. In process of time the network in Fig.~\ref{fig_hypernetwork}(a) can exhibit different configurations corresponding to particular hypersimplices in Fig.~\ref{fig_hypernetwork}(b), i.e., at each moment only one hypersimplex can be found in the network. All possible hypersimplices realized in the network form a simplicial family.

A structured set of hypercimplices determines a hypernetwork (Fig.~\ref{fig_hypernetwork}(c)) where the connections between hypercimplices describe their functional relations. In our model we assume that two hypercimplices are connected if the network can change its state from one hypersimplex to another due to the evolutionary rules governing its structure and dynamics. Internal as well as external conditions usually tend to slowly evolve or drastically change which causes the network to adapt by reconfiguring the structure of its links. In general this leads to rewiring and a new hypersimplex appears in the system.  Hence the co-evolution of structure and dynamics at the bottom-level network in Fig.~\ref{fig_hypernetwork}(a) gives rise to a kind of traffic at the upper level -- in the hypernetwork~(Fig.~\ref{fig_hypernetwork}(c)).  The complexity of the hypernetwork and potential transitions in it depend on the dynamical principles and evolutionary rules that govern the constituents of the basic network. An actual path in the hypernetwork depends on the joint action of external inputs and the network's internal state. Sensory information and task requirements activate some subset in the hypernetwork thus transforming it to a simpler structure that enables to move along the hypernetwork by a specific path. Each such path is a unique pattern of transient cluster activity for a given task performance and is robust in nonstationary environments.

\begin{figure}[h]
\centering\includegraphics[width=4.5in]{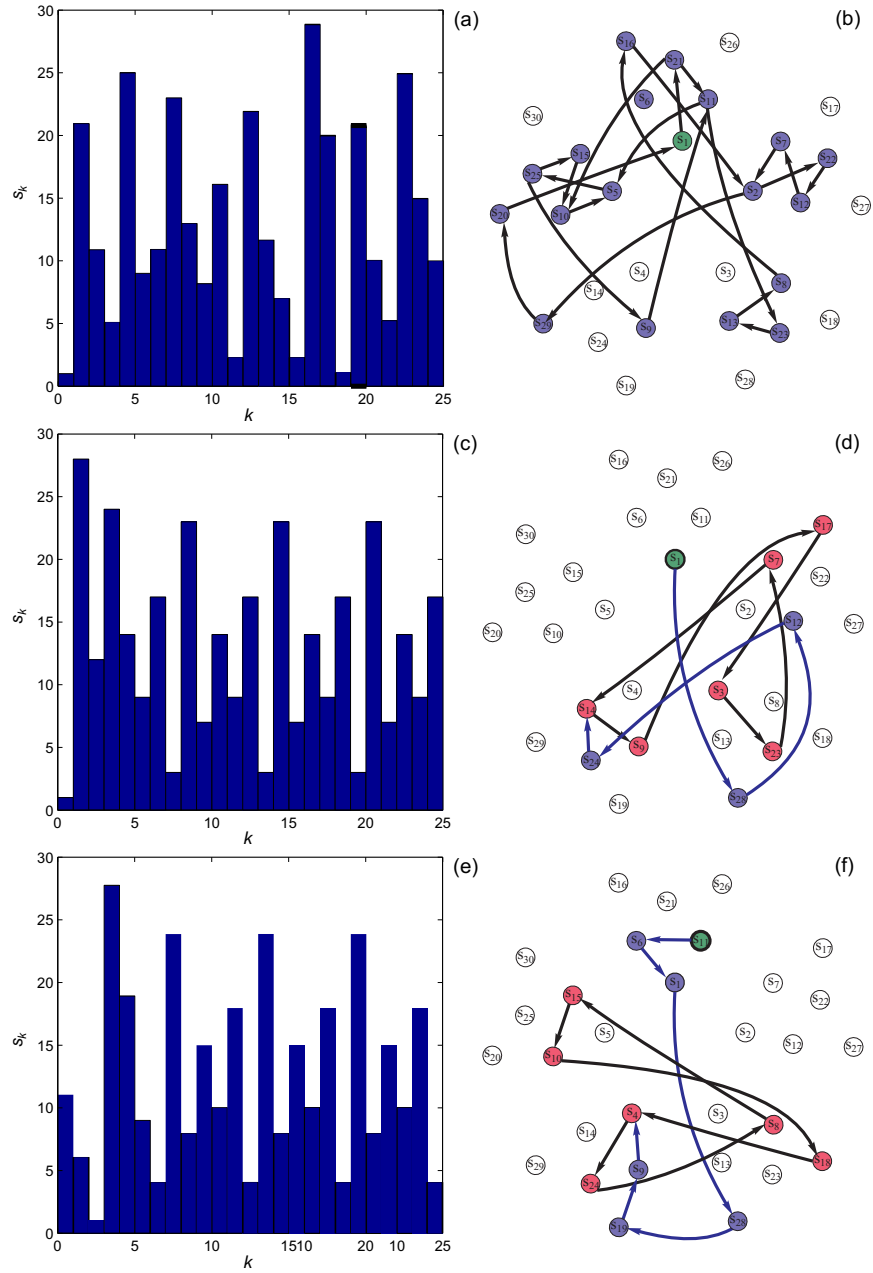}
\caption{The sequences of clusters states $\{s_k\}$ (a,c,e) and the corresponding paths in the hypernetwork (b,d,f) for the autonomous case (a,b) and under constant inputs (c-f). In the autonomous case one observes a random walk in the hypernetwork (b). In case of constant stimuli there are different scenarios: for example, when the stimulus is applied to the 1st node and the network is initially in the state $s_1$ (green circle) one obtains a path that after 4 steps comes to the 6-cycle of cluster states (red circles)  $s_{14}, s_{9}, s_{17}, s_{3}, s_{23}, s_{7}$; when the stimulus is applied to the 2nd node and the network is initially in the state $s_{11}$ there is a path that after 6 steps comes to the 6-cycle of cluster states  $s_{4}, s_{24}, s_{8}, s_{15}, s_{10}, s_{18}$. }
\label{fig_path}
\end{figure}

\subsection{Example: 5-node oscillatory network with inhibitory connections}
We illustrate this general paradigm by a model in the form of a 5-node dynamical network. The nodes-oscillators are coupled by inhibitory links in such a way that they can generate a rather simple but nontrivial form of activity: sequential clusters that appear cyclically. This is due to a chosen scheme of coupling where some two nodes inhibit the other two nodes, the latter inhibit the last node, and this one inhibits the former two nodes. The nodes-oscillators are tuned to have a property of post-inhibitory rebound: an oscillator receives an inhibitory input and then is released from it after which the oscillator fires a post-inhibitory burst of spikes. Inhibitory links sent by some oscillators to others allow to suppress the activity of the latter while the former are active. Thus the network can produce a 3-step cycle of clusters (which we refer to as a cluster state)  where first two nodes synchronously fire, then the other two generate synchronous bursts, after that  the remaining one becomes active, and finally it starts all over again. For example, at the cluster state $s_1$ nodes 1 and 2 fire synchronously first while the others are silent, then nodes 3 and 4 generate synchronous bursts, and finally node 5 fires while the other four are
below the threshold. Totally there are 30 cluster states $s_i$, $i=1\dots 30$, and corresponding topologies having the described properties (for more details, see \emph{Materials and Methods} and Ref.~\cite{maslennikov2015evolving}).

The coupling structure is static for some time interval, thus enabling to produce several identical cluster cycles. After that the structure switches to another one that depends on the dynamics exhibited by the oscillators. The new structure has the same symmetric properties as the previous one reflecting the homeostasis inherent to biological networks. In the new state the network also generates 3 cyclic clusters but they comprise other oscillators; in other words, the clusters mix during the structure switching by swapping the nodes between each other. We assume that the new topology depends on what cluster is active at the moment of switching; since we have 3 clusters there are 3 different topologies which can be established after the switching. From time to time the network moves from a current cluster state to one of three others.  The system is basically an adaptive network where the connection topology determines the form of nodal activity while the oscillatory dynamics influences the network structure affecting the subsequent coupling scheme after the switching. As we talked in the previous subsection, a structured set of nodes which determines a cyclic cluster sequence can be regarded as a  hypersimplex. Therefore we consider all 30 possible 3-cluster states in our network as a simplicial family. Being connected by directed links they form a hypernetwork of cluster states where some path is activated during the time course of the network co-evolution (see Fig.~\ref{fig_hypernetwork}).

\begin{figure}[h]
\centering\includegraphics[width=4.5in]{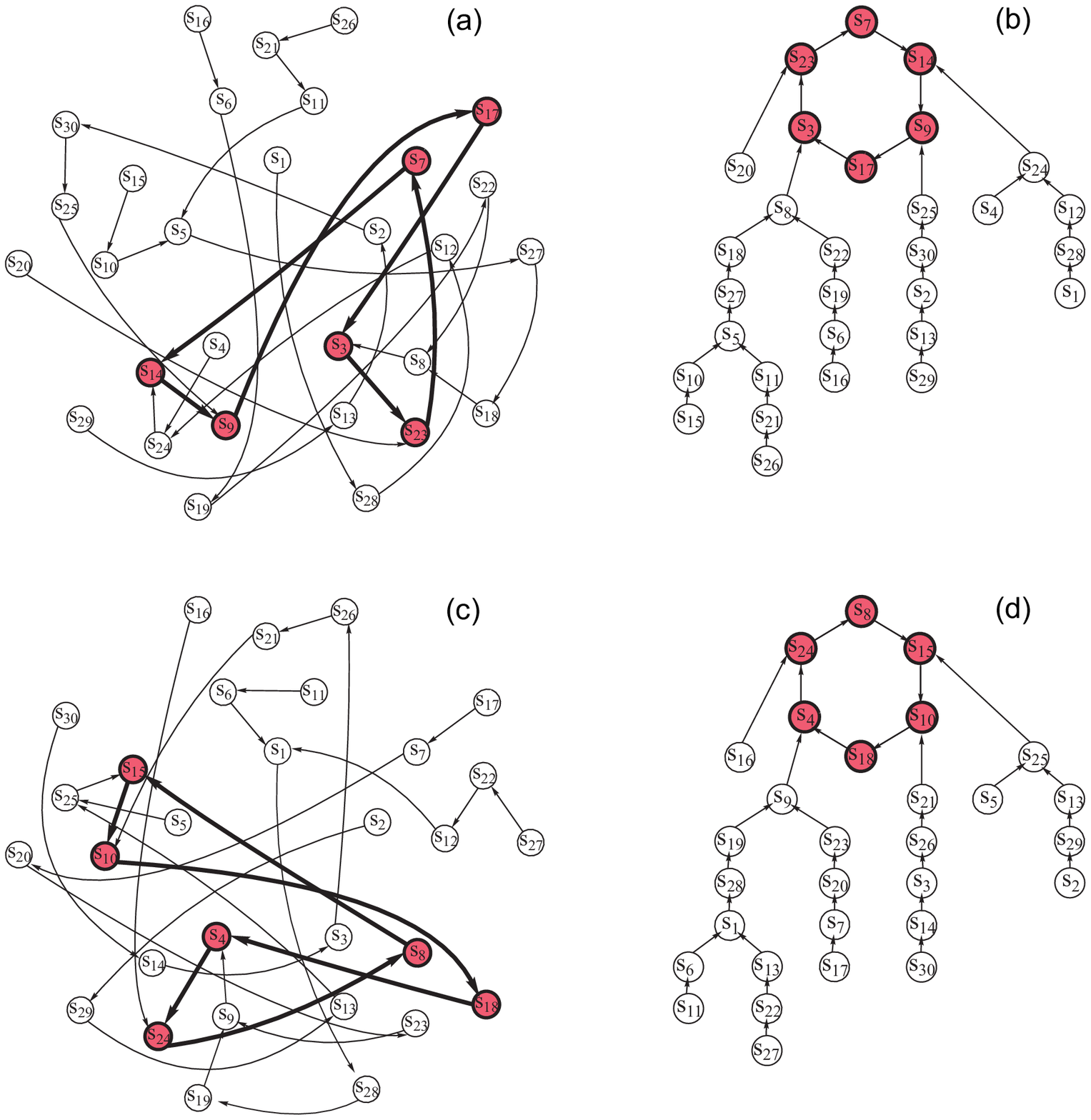}
\caption{The application of a constant stimulus to node~1~(a) and node~2~(c) specifies active links in the hypernetwork; thus this reduced hypernetwork governs transitions until  node~1 (respectively, node~2) remains under the stimulus. In both cases all the trajectories come to cycles of 6 cluster states shown by red circles. The same reduced hypernetworks can be presented in an untangled way (b,d). }
\label{fig_cycle}
\end{figure}

We studied the co-evolution of the network in an autonomous case and under the effect of constant stimuli. Left to itself, the network generates a sequence of different cluster states and the path in the corresponding hypernetwork is a random walk shown in Fig.~\ref{fig_path}(a,b). If an input is applied to the network then the path becomes stimulus-specified. The analysis of the network dynamics showed that the path depends on an internal network state -- its topology -- and the external stimulus, namely on what node it is applied to. In general, in the non-autonomous case the path in the hypernetwork after several steps comes to some cycle consisting of 6 cluster states and remains there until a new input comes. In the example shown in Fig.~\ref{fig_path}(c,d) the network initially is in the state $s_1$ and the constant input is applied to node 1. The trajectory in the hypernetwork goes through the sequence $s_1, s_{28}, s_{12}, s_{24}$ and comes to the cycle  $s_{14}, s_{9}, s_{17}, s_{3}, s_{23}, s_{7}$. When the input is applied to node 2 and the network is in the state $s_{11}$ at the beginning, the trajectory in the hypernetwork goes through the sequence $s_{11}, s_{6}, s_{1}, s_{28}, s_{19}, s_{9}$ and comes to the other cycle  $s_{4}, s_{24}, s_{8}, s_{15}, s_{10}, s_{18}$ shown in Fig.~\ref{fig_path}(e,f).

We found that the application of a constant stimulus to a network node specifies the subset of active directed links in the hypernetwork which govern the co-evolution while other links become inactive. In this new hypernetwork  each hypersimplex has only one outward link so the path becomes specified and there is a limiting path -- a 6-cycle -- attracting all other trajectories. For example, the application of a stimulus to node 1 result in a reduced hypernetwork shown in Fig.~\ref{fig_cycle}(a). Starting from some cluster state $s_i$, a trajectory moves along a well determined path towards a 6-cycle of states $s_{14}, s_{9}, s_{17}, s_{3}, s_{23}, s_{7}$. For a better visualization the same reduced hypernetwork is shown in Fig.~\ref{fig_cycle}(b) where it is seen how different branches of the reduced hypernetwork converge to the same 6-cycle. An input to node 2 leads to another reduced hypernetwork with a similar structure (Fig.~\ref{fig_cycle}(c,d)).

\begin{figure}[h]
\centering\includegraphics[width=5.0in]{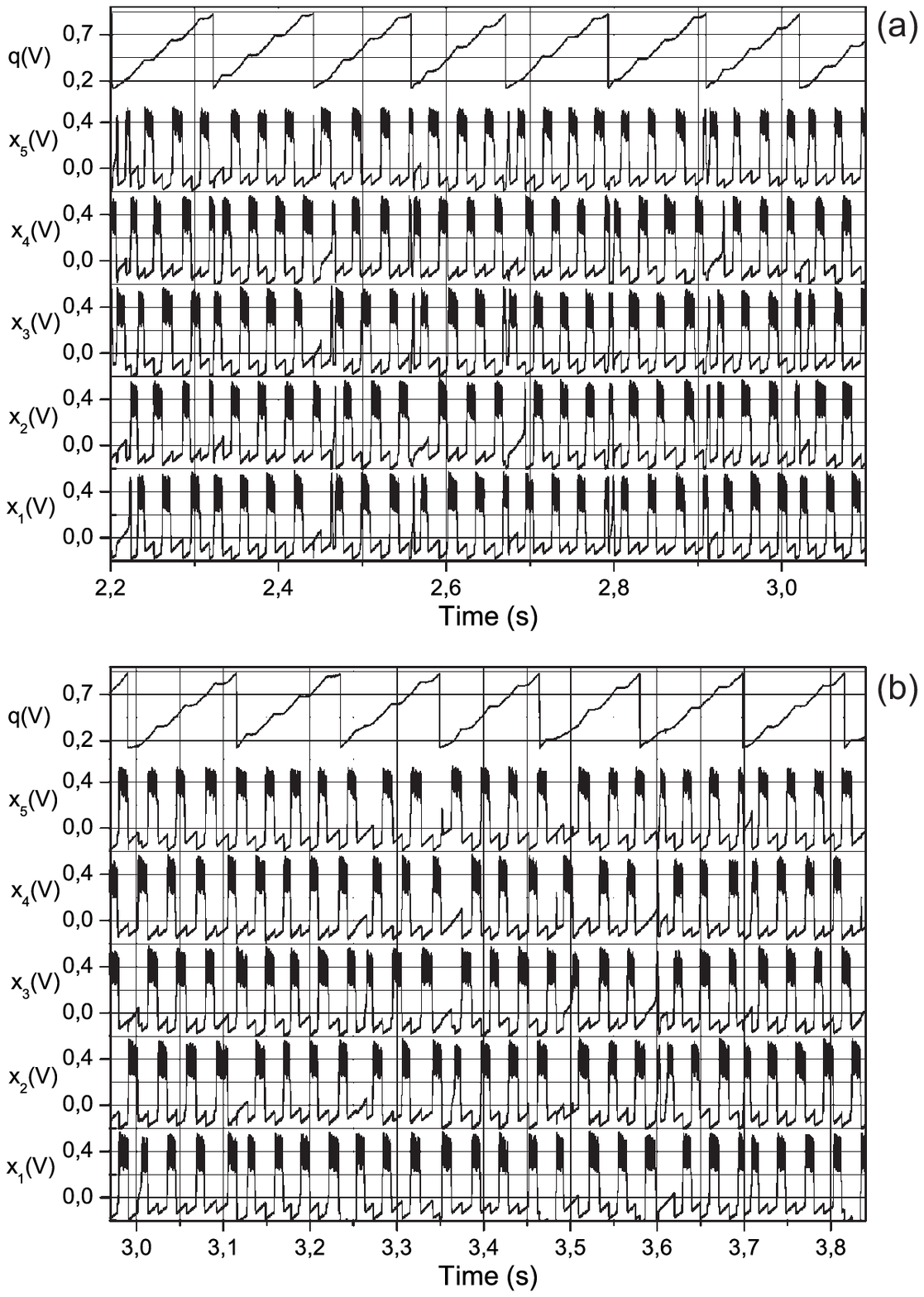}
\caption{Dynamics of the mean-field-based quantity $q$ and nodal variables $x_i$ ($i=1,\dots,5$) which represent transitions between different cluster states realized on FPGA Xilinx Artix-7. Time graphs are registered from multi-channel acquisition system. Waveforms in (a) correspond to a random walk in the hypernetwork in an autonomous case. Waveforms in (b) relate to the case of a stimulus-specific traffic in the hypernetwork (an input is applied to node 1) when the trajectory came to the 6-cycle $s_{23}, s_{7}, s_{14}, s_{9}, s_{17}, s_{3}, s_{23}$. }
\label{fig_experiment}
\end{figure}

To create a real-time functioning 5-node dynamical network with inhibitory connections we realized its electronic model on FPGA Xilinx Artix-7. The FPGA electronic model allows one to process external (sensory, informational) signals in this network in real time. Due to the real time processing and parameters scalability of the electronic model we have the opportunity to apply it to different types of interfaces (for example, a human-machine interface, a brain-computer interface, an artificial neural network interface, etc.). The analysis shows a full agreement of theoretical and numerical predictions with experimental data. Without external inputs the experimental setup  displays random transitions between different cluster states, an example is shown in Fig.~\ref{fig_experiment}(a). In case of an input application, the network demonstrates transitions relating to the corresponding reduced hypernetwork where each hypersimplex has only one outward link and all paths are converged to a 6-cycle. For example, when an input is applied to node 1 the network activity gives rise to a transition in the hypernetwork towards the cyclic sequence $s_{23}, s_{7}, s_{14}, s_{9}, s_{17}, s_{3}, s_{23}$, etc., shown in Fig.~\ref{fig_experiment}(b).

\section{Conclusion}
In this paper, we proposed a paradigmatic model of how a network of spiking neurons can create complex responses at the higher level or representation -- in the corresponding hypernetwork. Despite the simplicity of the coupling structure, the neuron model, and the evolutionary operator we show the basic idea: how dynamics of the adaptive oscillatory network of spiking neurons leads to the emergence of various transient behaviours in the hypernetwork. Consecutive switchings between different hypersimplices mean that each cluster state is unstable due to the evolutionary rules governing the system but applying inputs results in a robust stimulus-specific path in the hypernetwork. This is similar to dynamics exhibited by systems with the "winnerless competition" principle in Refs.~\cite{afraimovich2004origin,afraimovich2004heteroclinic,afraimovich2008winnerless,afraimovich2014hierarchical}, however the dynamic mechanism of the co-evolving topology and dynamics in our case is completely different than the stable heteroclinic channel in these works. Besides fundamental issues concerning the dynamic origin of relational simplices in large-scale neural networks there are several items relating to the case study model to be addressed in further research. For example, the influence of more complicated than constant inputs, the role of noise and parameters detuning on temporal properties of sequences in the hypernetwork (see, e.g.,~\cite{wordsworth2008spatiotemporal,ashwin2016mathematical}), the appearance of topology switching from local internodal interactions, and others.  These problems remain beyond the scope of this paper and provide a baseline for future studies.

\section{Materials and Methods}
\subsection{Nodal dynamics}
The activity of nodes-oscillators is described by the following  map~\cite{nekorkin2007diskretnaya,courbage2007chaotic}:
\begin{align}\label{eq:map2d}
\begin{split}
x_{i,n+1} &= x_{i,n} + F_H(x_{i,n}) - y_{i,n} + I_{i,n},\\
y_{i,n+1} &= y_{i,n} + \varepsilon(x_{i,n} - J_i), i=1,...,5,
\end{split}
\end{align}
where $n = 0,1,2,\dots$ is discrete time,  the variables $x_{i,n}$ and $y_{i,n}$ characterize the state of the $i$-th node at the moment of $n$. The nonlinear function $F_H(x) = x (x-a) (1-x) - \beta H(x-d)$, where $H(x)$ is the Heaviside step function: $H(x)=1$ if $x \geq 0$ and $H(x)=0$ otherwise; the parameters $a$, $\beta$, and $d$ control an oscillatory regime, in particular, $\beta$ and $d$  determine chaotic activity. The parameter $\varepsilon$ determines the rate for the variable $y_{i}$ and the parameter $J_i$ characterizes the excitatory properties of the nodes. The term $I_{i,n}$ takes into account the additive noise and the impact on the  $i$-th node from the other network nodes (inhibitory links);  the latter is given by
\begin{align}\label{eq:chem1}
\begin{split}
I_{i,n} = - g \sum\limits_{j=1, j\neq i}^{j=N} a_{ij, n} (x_{i,n}-\nu) H(x_{j,n}-\theta),
\end{split}
\end{align}
where the coefficient $g$ defines the coupling strength, $a_{ij, n}$ are the elements of an adjacency matrix $A=\{a_{ij}\}$ (see below) at the moment of $n$, $\nu$ is the so-called reversal parameter, and $\theta$ is the threshold parameter. In our simulations we fix the parameters $a=0.1$, $\beta=0.3$, $d=0.45$, $\varepsilon=10^{-3}$, $g=0.15$, $\nu=-0.5$, and $\theta=0.2$.  For these values, the nodes described by~\eqref{eq:map2d} and~\eqref{eq:chem1} display subthreshold activity in the absence of an input ($I_{i,n}=0$) and generate chaotic bursts as a result of the inhibition by the connected nodes (for more details, see~\cite{nekorkin2007diskretnaya,courbage2007chaotic}). The map-based approach in modeling neuronal networks is described in more detail in Refs.~\cite{courbage2010map,ibarz2011map,girardi2013brief,maslennikov2014map}. The system~\eqref{eq:map2d} was used in a number of research papers for modeling the collective activity of different neuronal systems (Refs.~\cite{nekorkin2011spike,courbage2012synchronization,maslennikov2012discrete,maslennikov2013emergence,maslennikov2014modular,maslennikov2015basin}) and studying rather theoretical issues on chaotic dynamics in maps (Refs.~\cite{maslennikov2013dynamic,maslennikov2016attractors}). Note that the chaos itself is not necessarily needed to generate a hypernetwork and the general results in Sec.~\ref{subsec_hypernetwork} do not depend on whether chaotic or regular dynamics is exhibited by the network. We use in our example chaotic oscillators to demonstrate the robustness of switching activity in adaptive networks of spiking neurons even in case of irregular dynamics.

\subsection{Dynamical network and cluster states}
We considered a network consisting of 5 nodes-oscillators.  The connections are directed and described by the adjacency matrix $A=\{a_{ij}\}$ where $a_{ij}=1$ if node $j$ sends a link to node $i$ and $a_{ij}=0$ otherwise. We consider the simplest nontrivial cluster states in this network, two clusters consist of two oscillators, and the remaining oscillator forms the third cluster. As a result, the network exhibits a cyclic sequence $\langle(i_1, i_2), (i_3, i_4), i_5\rangle$ which means that first two nodes with numbers $(i_1, i_2)$  fire synchronously (the first cluster), then nodes  $(i_3, i_4)$ generate synchronous bursts, after that node $i_5$ fires and the sequence repeats. To obtain this configuration we choose the elements $a_{ij}$ in a specific way. We take $a_{i_1i_5}=a_{i_2i_5}=1$, $a_{i_3i_1}=a_{i_3i_2}=a_{i_4i_1}=a_{i_4i_2}=1$, $a_{i_5i_3}=a_{i_5i_4}=1$, and the other elements of $A$ are equal to 0. The complete list of possible 3-cluster sequences of this configuration is presented in Table~\ref{tab:states} where each column is generated by cyclic permutation of the indexes $\langle(i_1, i_2), (i_3, i_4), i_5\rangle$ in the top state.

\vspace*{-7pt}
\begin{table}[!h]
\caption{List of the cluster states in the 5-node network}
\label{tab:states}
\begin{tabular}{lllll}
\hline
$s_{1}  =  \langle(1,2), (3,4), 5\rangle$ & $s_{6}  =  \langle(1,3), (2,4), 5\rangle$ & $s_{11} = \langle(1,4), (2,3), 5\rangle$  \\
$s_{2}  = \langle(2,3), (4, 5), 1\rangle$ & $s_{7}  = \langle(2,4), (3,5), 1\rangle$  & $s_{12} = \langle(2,5), (3,4), 1\rangle$  \\
$s_{3}  = \langle(3,4), (5,1), 2\rangle$  & $s_{8}  = \langle(3,5), (4,1), 2\rangle$  & $s_{13} = \langle(3,1), (4,5), 2\rangle$  \\
$s_{4}  = \langle(4,5), (1,2), 3\rangle$  & $s_{9}  = \langle(4,1), (5,2), 3\rangle$  & $s_{14} = \langle(4,2), (5,1), 3\rangle$  \\
$s_{5}  = \langle(5,1), (2,3), 4\rangle$  & $s_{10} = \langle(5,2), (3,1), 4\rangle$  & $s_{15} = \langle(5,3), (1,2), 4\rangle$  \\
\hline\\
\hline
$s_{16} = \langle(2,3), (1,4), 5\rangle$  & $s_{21} = \langle(2,4), (1,3), 5\rangle$ & $s_{26} = \langle(3,4), (1,2), 5\rangle$   \\
$s_{17} = \langle(3,4), (2,5), 1\rangle$  & $s_{22} = \langle(3,5), (2,4), 1\rangle$ & $s_{27} = \langle(4,5), (2,3), 1\rangle$ \\
$s_{18} = \langle(4,5), (3,1), 2\rangle$ & $s_{23} = \langle(4,1), (3,5), 2\rangle$ & $s_{28} = \langle(5,1), (3,4), 2\rangle$ \\
$s_{19} = \langle(5,1), (4,2), 3\rangle$ & $s_{24} = \langle(5,2), (4,1), 3\rangle$ & $s_{29} = \langle(1,2), (4,5), 3\rangle$\\
$s_{20} = \langle(1,2), (5,3), 4\rangle$ & $s_{25} = \langle(3,1), (5,2), 4\rangle$ & $s_{30} = \langle(2,3), (5,1), 4\rangle$ \\
\hline
\end{tabular}
\vspace*{-4pt}
\end{table}

\subsection{Co-evolution of nodal dynamics and coupling topology} The coupling topology evidentally influences the nodal dynamics since it directly determines the form and order of clusters. To take into account the impact of the dynamics on the coupling topology we use the following algorithm. We introduce an auxiliary variable $q$ which defines the moment $n=n^*$ of topology switching in the following way:
\begin{align}\label{eq:q}
\begin{split}
q_{n+1} &= q_{n} + \mu X_n, \quad X_n = \frac{1}{5} \sum\limits_{i=1}^5 x_{i,n},  \\
\text{if} \; q_n &>1 \,  \text{then} \, q_{n}:=0 \, \text{and}\, n^*:=n. \\
\end{split}
\end{align}
It follows from Eqs.~\eqref{eq:q} that $q$ is increasing on the average starting from zero up to the threshold $q=1$ with the rate determined by the network behavior (the mean field $X$) and a small parameter $\mu$ ($0<\mu\ll1$). The dynamics of $X$ is chaotic due to the chaotic activity of  $x_{i}$, $i=1,\dots,5$, which leads to small perturbations of $q$. After reaching the threshold, the value of $q$ is reset to zero and begins to grow according to Eq.~\eqref{eq:q}. The topology rewiring occurs at the moment  $n=n^*$ of resetting of $q$.

At the moment of rewiring, two nodes are chosen, one from the current active cluster, and another from the previous active cluster. The indexes of these nodes, $k$ and $l$, determine how the coupling topology is changed according to the rule:
\begin{align}\label{eq:TAT}
\begin{split}
A_{n+1} = T_{kl} A_{n} T_{kl}.
\end{split}
\end{align}
Which numbers $k$ and $l$ are chosen depends on the following algorithm. Suppose at the moment of rewiring $n=n^*$ the nodes with indexes $(i_3, i_4)$  are active, and before that the nodes  $(i_1, i_2)$ were active. In the clockwise ordered set starting from $i_5$  there is a pair of nodes $i_{k^*}\in(i_3, i_4)$ and $i_{l^*}\in(i_1, i_2)$  that have a minimum index distance determined from the clockwise ordering of the nodes. For example,  the index clockwise distance between nodes 2 and 3 is 1 while between 3 and 2 is 4.  The indexes $i_{k^*}$ and $i_{l^*}$ specify the numbers $k = k^*$ and  $l = l^*$ in Eq.~\eqref{eq:TAT}.
The matrix $T_{kl}$ is obtained by swapping row $k$ and row $l$ of the identity $5\times5$ matrix with 1's on the main diagonal and 0's elsewhere. Thus $T_{kl}$ makes a row-switching transformation, i.e., $T_{kl} A$ is the matrix produced by exchanging row $k$ and row $l$ of $A$. The product $T_{kl} A T_{kl}$ is thus a matrix where firstly the rows $k$ and  $l$ are exchanged and then in the obtained matrix the columns $k$ and $l$ are swapped.

\subsection{Experimental setup}
The discrete-time systems describing the nodal and coupling dynamics in Eqs.~\eqref{eq:map2d}~and~\eqref{eq:chem1} as well as the evolutionary rule governing the topology switching in Eqs.~\eqref{eq:q}~and~\eqref{eq:TAT} were realized on FPGA Xilinx Artix-7.  The signal from each oscillator of the network and the signal corresponding to average variable $q$ were registered from the FPGA through the 12-bit digital-to-analog convertors. One discrete time step for this network is equal 50 $\mu s$, the average spike duration is about 10-20 discrete steps (about 0.5-1 ms), and the burst lasts  about 25 ms. Such timings are well correspond to time scales of real neurons. The parameters are equal to $\mu=0.001$, $\varepsilon=0.001$, $J=0.05$, $\beta=0.3$, $a=0.1$, $d=0.45$, $\theta=0.2$, $\nu=-0.5$, and $g=0.07$.

%{O.V.M. performed simulations and drafted the manuscript. D.S.S. carried out the experiments. V.I.N. conceived of and designed the study, and drafted the manuscript. All authors read and approved the manuscript.}

%{The authors declare that they have no competing interests.}
\section*{Acknowledgement}
{This work was supported by the Russian Science Foundation (Project No. 16-42-01043).}

\end{document}